\documentclass[%
reprint,
superscriptaddress,
 amsmath,amssymb,
 aps,
pra,
floatfix,
]{revtex4-1}
\usepackage{graphicx}
\usepackage{dcolumn}
\usepackage{bm}
\usepackage{hyperref}
\usepackage{gensymb}

\begin{document}

\preprint{APS/123-QED}

\title{Controlling nonlinear rogue wave formation using the coherence length of phase noise}
\author{Saumya Choudhary}
\email{schoudha@ur.rochester.edu}
\affiliation{Institute of Optics, University of Rochester, Rochester, NY 14627, USA}%
\author{A. Nicholas Black}
\affiliation{Department of Physics and Astronomy, University of Rochester, Rochester, NY 14627, USA}
\author{Aku Antikainen}
\affiliation{Institute of Optics, University of Rochester, Rochester, NY 14627, USA}
\author{Robert W. Boyd}
\affiliation{Institute of Optics, University of Rochester, Rochester, NY 14627, USA}%
\affiliation{Department of Physics and Astronomy, University of Rochester, Rochester, NY 14627, USA}
\affiliation{Department of Physics, University of Ottawa, Ottawa, Ontario, K1N 6N5, Canada}%

\begin{abstract}
Weak phase noise present on an optical field can be amplified by a self-focusing nonlinearity and form intense ``rogue wave" features. Here, we study the effect of the coherence length (or grain size) of this phase noise on the likelihood of rogue wave formation in the presence of a self-focusing nonlinearity. We show that while the likelihood of rogue wave formation increases with laser power when the coherence length is only slightly smaller that the beam diameter, the likelihood is minimally affected by change in laser power when the coherence length is significantly smaller than the beam diameter. Our study provides insight into the interaction of nonlinearity with phase instabilities on a field, and could be useful in applications such as reducing the effect of turbulence-induced breakup of intense laser beams, and developing radiance limiters to reduce the focusable power in a beam.
\end{abstract}

\maketitle

The formation of rare but extreme (or ``rogue") amplitude waves in optical  \cite{Solli2007, Liu2015, Mathis2015, Pierangeli2015, Pierangeli2016, Safari2017a}, microwave \cite{Hohmann2010}, and hydrodynamic systems \cite{Degueldre2016} have attracted considerable recent interest \cite{Onorato2013, Dudley2014, Dudley2019}. These rogue waves have a non-Gaussian probability distribution of the wave amplitude with a long-tailed probability distribution of the intensity. A random phase fluctuation with sufficient strength imposed on a beam can develop on propagation into network-like intensity patterns that are commonly referred to as ``caustics" \cite{Mathis2015, Safari2017a}. Light can concentrate very tightly in caustics, which facilitates rogue wave formation and leads to long-tailed intensity statistics. Rogue waves in linear systems can develop through the constructive interference of several waves with random phases and amplitudes \cite{Longuet1957}, or through the directional focusing of these waves \cite{Fochesato2007}. Speckle formation in optical systems is also a linear phenomenon, and a fully developed speckle pattern has Gaussian statistics in the wave amplitude \cite{Goodman1976}. Non-Gaussian amplitude statistics in linear systems can also occur due to multiple scattering through a medium \cite{Nieuwenhuizen1995}, due to the spatial inhomogeneity-induced clustering of speckles with different grain sizes \cite{Arecchi2011}, and through the redistribution of energy among several speckle grains due to higher-order correlations encoded onto the field \cite{Bromberg2014}. 

The presence of nonlinearity in an optical system can considerably influence the formation of rogue waves. Rogue events have been observed during supercontinuum generation in nonlinear fiber-optics systems and are the result of collisions between ``breather" solitons formed by nonlinear amplification of modulational instability in the system \cite{Solli2007, Dudley2014, Walczak2015, Narhi2016}. Rogue waves can also form in spatially extended nonlinear systems either by means of self-focusing seeded by wavefront perturbations on the field \cite{Onorato2006, Safari2017a, Pierangeli2015} or by hypercycle amplification after the breaking of spatial symmetry in optical cavities \cite{Montina2009}. Small scale filamentation is another phenomenon that occurs when a large self-focusing nonlinearity amplifies angular spectral sidebands through four-wave mixing, leading to the formation of several localized structures called ``filaments" such that each filament has the same (critical) power $P_{\rm cr}$ \cite{Boyd_NLO, Birkholz2013}. Rogue waves can also form in a beam undergoing small-scale filamentation when filaments merge because of medium inhomogeneities \cite{Birkholz2013}. A self-focusing nonlinearity can enhance rogue wave formation in laser beams containing weak phase noise \cite{Safari2017a}. However, a non-uniform polarization structure on the beam can suppress rogue waves under certain conditions \cite{Black2022}. Rogue waves are more likely to form in speckle patterns of a particular coherence length propagating through a photorefractive crystal due to the saturation of nonlinearity once a rogue feature reaches a certain minimum width \cite{Pierangeli2016}. Light scattered through a medium with tailored disorder can also show a similar enhancement of rogue wave formation at a particular coherence length of the disorder \cite{Zannotti2019}. 

Here, we study how the transverse spatial coherence length of phase noise affects rogue wave formation in the presence of a self-focusing nonlinearity. We measure the intensity statistics of the beam  after it propagates through a hot rubidium vapor cell for various coherence lengths (or grain sizes) of the phase noise and various beam powers. We observe that the intensity statistics have a diminished sensitivity to nonlinearity when the coherence length of the phase noise is much smaller than the beam width. We also study the mechanism behind this effect through numerical simulations of nonlinear beam propagation. Our simulations show that small-grained phase noise induces the redistribution of beam power into multiple filaments of reduced intensity, thereby limiting the maximum intensity in a rogue feature relative to the background. Our study complements Refs. \cite{Pierangeli2016, Safari2017a, Black2022}, and is relevant for the development of better optical power limiters, and for probing a turbulent medium and mitigating its effect on the propagation of intense laser beams.

\section{Experiment}

\begin{figure}[ht!]
	\includegraphics[width=0.48\textwidth]{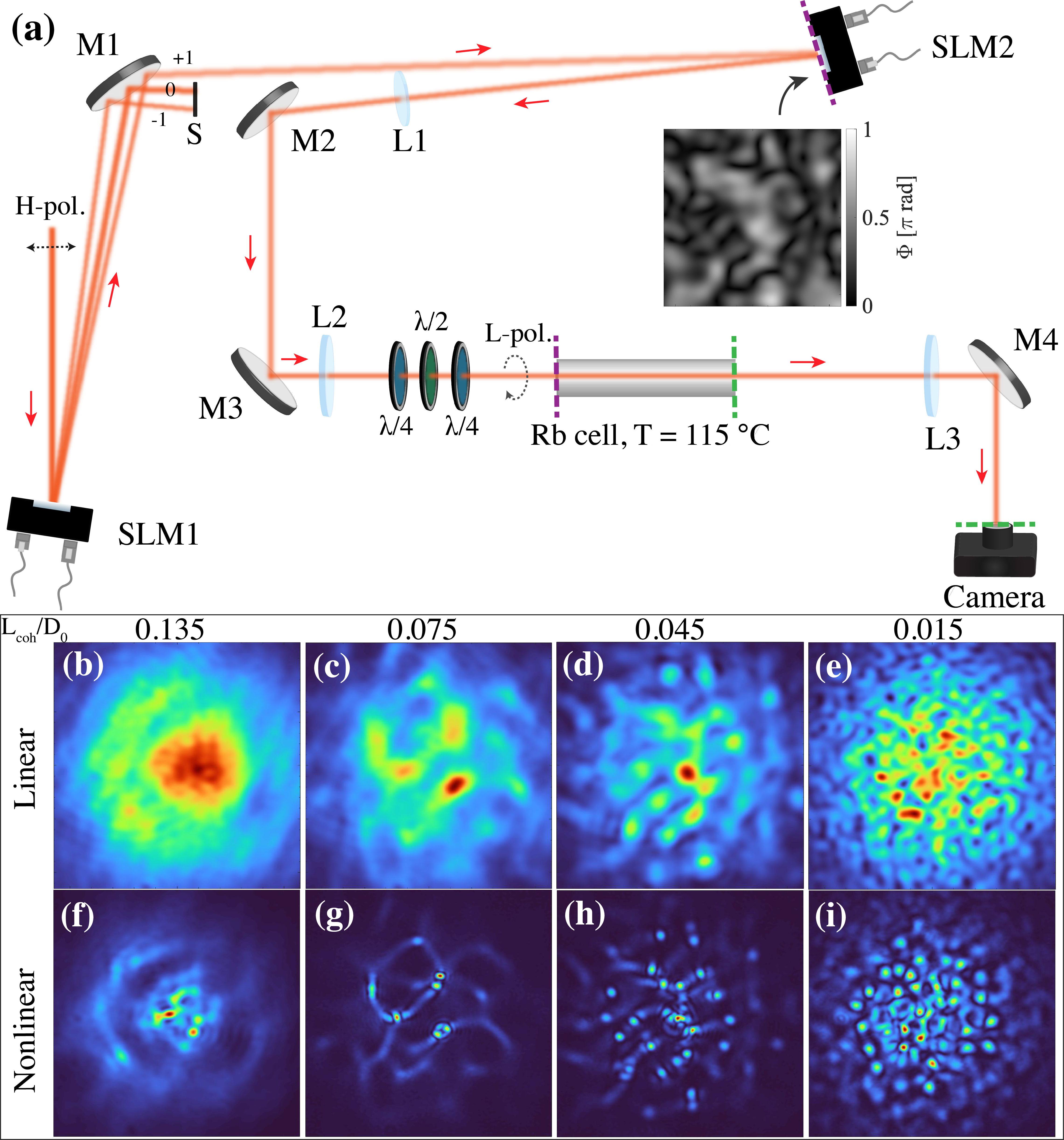}
	\caption{(a) Schematic of the experimental setup. H-polarized light from a tunable diode laser is diffracted by a phase-only grating on spatial light modulator SLM1 forming a Gaussian beam of diameter $D_0$ (to $1/e^2$ values of the intensity) in the first diffractive order. SLM2 adds a random phase mask (example shown in the inset) with coherence length $L_{\rm coh}$ and amplitude of $\pi$ rad on the beam. SLM2 is then imaged using lenses L1 and L2 onto the entrance facet (purple dashed line) of a 7.5 cm-long rubidium cell heated to 115 \degree C . The output facet (green dashed line) of the cell is then imaged by lens L3 onto the image plane of the camera. Measured caustic patterns formed by noisy beams with $L_{\rm coh}/D_0$ of (b, f) 0.135, (c, g) 0.075, (d, h) 0.045, and (e, i) 0.015, after linear (top), and nonlinear (bottom) propagation through the cell are also shown. The beam power at the input of the cell ($P_{\rm in}$) was 90 mW for the nonlinear results. The focal lengths of the lenses L1, L2 and L3 are 1 m, 75 cm and 30 cm, respectively. }
	\label{fig:fig1}
\end{figure}

Figure~\ref{fig:fig1}(a) shows the schematic of our experimental setup. Our saturable nonlinear medium is a cell containing natural abundance rubidium. We heat the cell to 115 \degree C, and blue detune our laser source by 600 MHz above the $^{87}$Rb $D_2$ $F = 1 \rightarrow F' = 2$ transition in order to have a self-focusing nonlinear response. A horizontally polarized beam from our laser source diffracts from a phase grating impressed on a spatial light modulator (SLM1) and forms a Gaussian beam of diameter 2.5 mm ($D_0$) in the first diffractive order. We isolate this diffractive order by letting the light propagate over 2 m, and add a conjugate defocus on SLM1 to compensate for the accumulated defocus on the beam. Both SLM1 and SLM2 are liquid-crystal-on-silicon (LCOS) phase only SLMs from Hamamatsu that have identical resolution ($600 \times 800$) and pixel size (20 $\mu$m). SLM2 adds a random phase mask with a spatial coherence length $L_{\rm coh}$ and a maximum amplitude of $\pi$ rad onto the beam. To determine the random phase mask, we generate a $600 \times 800$ matrix of uniformly distributed random numbers between 0 and 1, and convolve it with a Gaussian point spread function of width $L_{\rm coh}$ [Eq.~(\ref{eq:eqB1}) in appendix B], which acts as a blur on the salt-and-pepper noise matrix \cite{Shirai2005}. We then multiply the matrix by $\pi$ so that the maximum phase amplitude of the added phase noise is $\pi$ rad. Limiting the maximum phase amplitude to $\pi$ rad ensures, as we show later, that rogue waves do not develop after purely linear propagation of the beam through the rubidium cell. The lenses L1 and L2 image the active area of SLM2 onto the entrance facet of the rubidium cell. The waveplates before the cell convert the polarization of the beam to left-handed circular to match the handedness of the $\sigma_+$ atomic transition. The lens L3 images the output facet of the cell onto the image plane of the camera, which records the intensity at the cell output.

Figures~\ref{fig:fig1}(b)-(e) show the recorded output intensity distributions after linear propagation through the cell for representative phase masks with $L_{\rm coh}/D_0$ of 0.135, 0.075, 0.045, and 0.015, respectively. For all linear measurements, we increase the value of detuning from 600 MHz to 65.04 GHz and fix input beam power $P_{\rm in}$ to 4 mW. As shown in Figs.~\ref{fig:fig1}(b)-(e), the added phase noise leads to distortion of the beam intensity upon linear propagation, but is weak enough that no sharp caustics are formed. As we decrease the $L_{\rm coh}/D_0$ of noise (left to right), more ``hotspots" are formed in the beam such that the intensity corresponding to the smallest $L_{\rm coh}/D_0$ [Fig.~\ref{fig:fig1}(e)] starts to resemble a speckle pattern. Figures~\ref{fig:fig1}(f)-(i) show the recorded intensities for the same phase masks as in the top panels (b)-(e), but with the nonlinearity turned on by changing the detuning to 600 MHz, and the beam power $P_{\rm in}$ to 90 mW. The nonlinearity sharpens the hotspots formed during linear propagation while preserving their underlying structure \cite{Safari2017a}. 

\begin{figure}[ht!]
	\centering
	\includegraphics[width=0.36\textwidth]{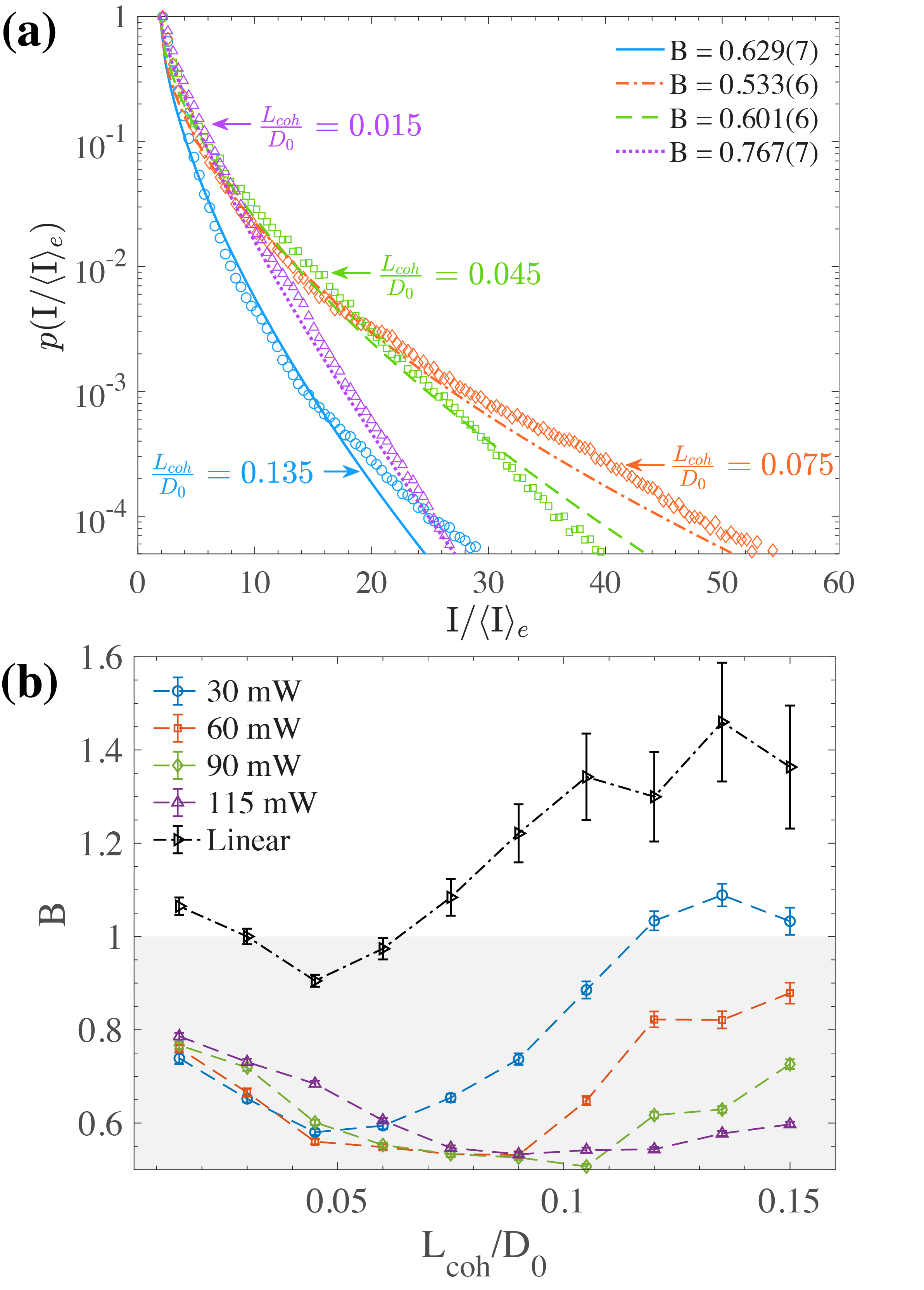}
	\caption{(a) Measured intensity distributions (markers), and their respective stretched exponential fits (lines) for $P_{\rm in}$ of 90 mW, and $L_{\rm coh}/D_0$ of 0.135 (blue, solid), 0.075 (red, dot-dashed), 0.045 (green, dashed), and 0.015 (purple, dotted). The value of the intensity exponent $B$ for each fit is stated in the legend. (b) The variation of $B$ with $L_{\rm coh}/D_0$ for linear measurements (black triangles), and nonlinear measurements with $P_{\rm in}$ of 30 mW (blue circles), 60 mW (red squares), 90 mW (green diamonds), and 115 mW (purple triangles). The gray shaded region indicates the range of $B$ corresponding to long-tailed intensity statistics.}
	\label{fig:fig2}
\end{figure}

To quantify the intensity statistics, we record output intensity patterns for an ensemble of 500 random phase masks with the same $L_{\rm coh}$. We acquire these intensity datasets for nonlinear propagation through the cell at various incident beam powers $P_{\rm in}$ (30 mW, 60 mW, 90 mW, and 115 mW) and various $L_{\rm coh}$ values (varied from 50 $\mu$m to 450 $\mu$m). We also record datasets for linear propagation through the cell. The probability distribution of intensities $p(I/ \langle I \rangle_e)$ in these datasets is well described by the following stretched exponential distribution \cite{Pierangeli2015, Safari2017a, Black2022} 
\begin{equation}
    p\Big( \frac{I}{\langle I \rangle}_e \Big) = N \exp \Big[ - A \Big( \frac{I}{\langle I \rangle}_e \Big)^B \Big].
    \label{eq:eq1}
\end{equation}
Here, $\langle I \rangle_e$ is the intensity averaged over the entire dataset, $N$ is the normalization coefficient, $A$ describes the width of the distribution, and $B$ is the stretching coefficient that determines its tail. When $B = 1$, the distribution is the usual exponential function associated with a fully developed speckle \cite{Goodman1976}. As $B$ becomes smaller than 1, caustic formation and rogue wave behavior are more likely to occur, and the intensity statistics become more long tailed. To estimate $B$ for each dataset, we fit Eq.~(\ref{eq:eq1}) to the tails of the respective intensity histograms using maximum likelihood estimation (MLE) and use Monte-Carlo simulations to obtain the uncertainties of the fit parameters. 

Figure~\ref{fig:fig2}(a) shows the measured intensity statistics along with their respective MLE fit for $P_{\rm in}$ of 90 mW, and $L_{\rm coh}/D_0$ of 0.135 (blue circles and solid line), 0.075 (red diamonds and dot-dashed line), 0.045 (green squares and dashed line), and 0.015 (purple triangles and dotted line). We note that phase noise of smaller $L_{\rm coh}/D_0$ has a wider angular spectral bandwidth [see Fig.~\ref{fig:figB} in appendix B]. This broadband noise seed should cause further broadening of the angular spectrum of the beam through four-wave mixing and lead to sharper caustics and longer-tailed intensity statistics. However, we do not observe a monotonic increase in the `tailiness' of intensity statistics as $L_{\rm coh}/D_0$ is reduced in Fig.~\ref{fig:fig2}(a), which is also reflected in the associated values of $B$ given in the legend. Instead, $B$ is minimized for $L_{\rm coh}/D_0$ of 0.075, and its distribution is the most long tailed.

Figure~\ref{fig:fig2}(b) shows the variation of $B$ with $L_{\rm coh}/D_0$ for linear measurements (black triangles) and nonlinear measurements with $P_{\rm in}$ of 30 mW (blue circles), 60 mW (red squares), 90 mW (green diamonds), and 115 mW (purple triangles). The shaded gray region represents the range of $B$ for which rogue wave behavior is likely. For linear measurements, we observe that $B$ is larger than 1 for most $L_{\rm coh}/D_0$, and we do not see evidence of either caustic or speckle formation. This result is in agreement with the fact that our maximum phase amplitude is $\pi$ rad and hence too low to form either linear caustics \cite{Safari2017a} or a fully developed speckle for which the total phase excursion by the scattering induced random walk should be at least $2\pi$ rad \cite{Goodman1976, Goodman2007}. The value of $B$ for nonlinear measurements is smaller than $B$ for linear measurements for all $L_{\rm coh}/D_0$, which is consistent with the aforementioned increase in sharpness of caustics due to nonlinearity. The noteworthy feature, however, is that for nonlinear measurements, $B$ is significantly more sensitive to the beam power $P_{\rm in}$ when $L_{\rm coh}/D_0$ is larger than 0.075 than it is for smaller $L_{\rm coh}/D_0$. We further explore this reduced sensitivity to nonlinearity of rogue wave formation for beams with broadband phase noise through numerical simulations.

\section{Numerical Modeling}

\begin{figure}[ht!]
	\centering
	\includegraphics[width=0.35\textwidth]{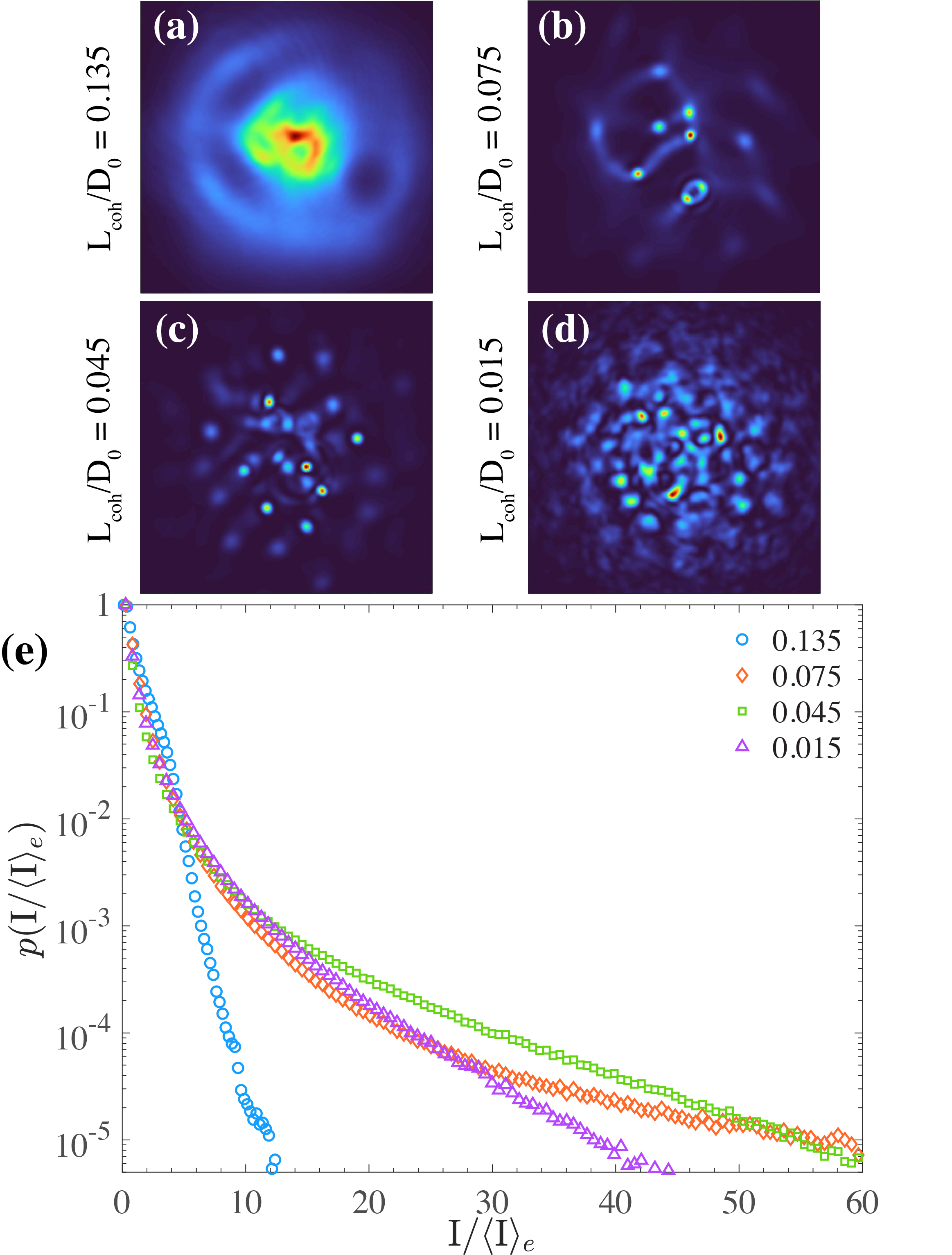}
	\caption{(a) Simulated caustic patterns at the output of the cell for $P_{\rm in}$ of 90 mW, and $L_{\rm coh}/D_0$ of (a) 0.135, (b) 0.075, (c) 0.045, and (d) 0.015. The phase masks used for these calculations were the same as the ones used in the experiment to capture the caustic patterns shown in Figs.~\ref{fig:fig1}(b)-(e). (b) Simulated intensity statistics for $P_{\rm in}$ of 90 mW, and $L_{\rm coh}/D_0$ of 0.135 (blue circles), 0.075 (red diamonds), 0.045 (green squares), and 0.015 (purple triangles).}
	\label{fig:fig3}
\end{figure}

The propagation of a field $\boldsymbol{E}(\boldsymbol{r}, t) = E(x, y) e^{i(kz - \omega t)} \hat{\boldsymbol{e}_L} + \text{c.c.}$ through a spatially extended nonlinear medium, such as our rubidium cell, can be described by the (2+1)-D nonlinear Schr\"odinger equation (NLSE) \cite{Boyd_NLO} given below
\begin{equation}
    \frac{\partial E}{\partial z} - \frac{i}{2k} \nabla^2_{\perp} E = \frac{ik}{2\epsilon_0}P,
    \label{eq:eq2}
\end{equation}
where $E(x, y)$ is the field envelope,  $\omega$ is the angular frequency of the laser, $k$ is the wave number, $\nabla^2_{\perp} = \partial^2/\partial x^2 + \partial^2/\partial y^2$ is the transverse Laplacian, $P = \epsilon_0 \chi(E) E$ is the atomic polarization, and $\chi (E)$ is the total atomic susceptibility that includes the linear as well as all orders of nonlinear response \cite{Boyd_NLO}. In our calculation of total susceptibility, we include the contributions from all the D2 transitions of rubidium. See appendix A for more details. We use the split-step Fourier method \cite{Agrawal2013} to solve Eq.~(\ref{eq:eq2}), and obtain the field at any location $(x, y, z)$ within the rubidium cell. We use Fresnel propagation for all linear propagation calculations \cite{GoodmanFourier}. For all simulations, we assume a transverse resolution of 2048 $\times$ 2048 pixels, a pixel size of 4.89 $\mu$m, and a longitudinal step size of 0.5 mm.  

Figures~\ref{fig:fig3}(a)-(d) show the simulated output intensities for the same set of phase masks used in the experiment that were used for the measured output intensities shown in Figs.~\ref{fig:fig1}(f)-(i). We also include an amplitude mask on the Gaussian beam to match the intensity of the Gaussian beam in our experiment [see Fig.~\ref{fig:figC1}(a)]. The simulated intensities in Figs.~\ref{fig:fig3}(a)-(d), and the measured intensities in Figs.~\ref{fig:fig1}(f)-(i) have very similar underlying intensity structures and sharpness of caustic features. Figure~\ref{fig:fig3}(e) shows the simulated intensity statistics for $P_{\rm in}$ of 90 mW, and $L_{\rm coh}/D_0$ of 0.135 (blue circles), 0.075 (red diamonds), 0.045 (green squares), and 0.015 (purple triangles). We use 200 realizations of random phase masks of a particular $L_{\rm coh}/D_0$ to calculate these intensity statistics. The simulated statistics show a good qualitative agreement with the measured statistics shown in Fig.~\ref{fig:fig2}(a) for the same set of parameters, and in both scenarios, the histogram corresponding to $L_{\rm coh}/D_0$ of 0.075 is the most long tailed. There are several contributing factors in the experiment that could lead to the observed differences between simulations and measurements, such as nonlocality in the nonlinear response of rubidium vapor \cite{Suter1993}, aberrations in the imaging optics and the windows of the cell, and the pixel size of SLMs. However, our simplified numerical model agrees reasonably well with the measurements, and can be used to study the propagation dynamics of caustic and rogue wave formation within the cell. 

\section{Discussion}
We use the scintillation index $\beta^2$ as a metric for the sharpness of caustics, which is defined as follows \cite{Gbur2014, Kravtsov2012} 
\begin{equation}
    \beta^2 = \frac{\langle I^2 \rangle - \langle I \rangle^2}{\langle I \rangle^2}.
\end{equation}
\begin{widetext}

\begin{figure}[ht!]
	\centering
	\includegraphics[width=0.95\textwidth]{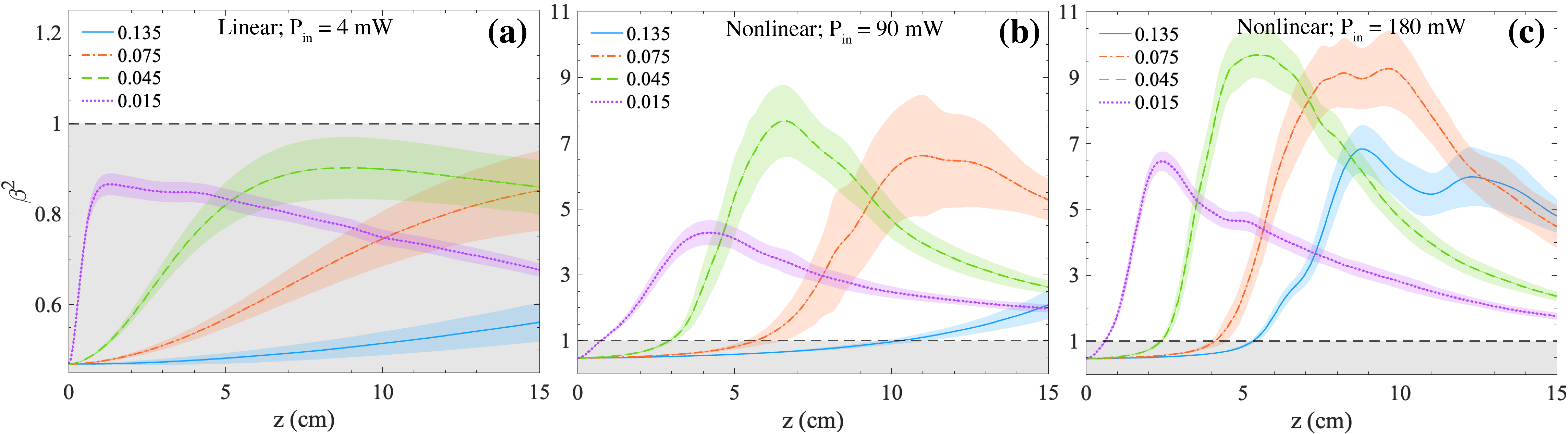}
	\caption{The evolution of scintillation index $\beta^2$ with propagation distance $z$ as predicted by our numerical model under (a) linear, and under (b), (c) nonlinear propagation with $P_{\rm in}$ of 90 mW and 180 mW, respectively. The legend shows the values of $L_{\rm coh}/D_0$ of the added random phase masks on the beam. The dashed black line indicates the threshold above which long-tailed intensity statistics start to emerge. }
	\label{fig:fig4}
\end{figure}

\end{widetext} 
Here, $\langle\cdots\rangle$ denotes the transverse spatial average over the entire field. For fully developed speckle patterns, $\beta^2$ is unity. In contrast, caustic patterns with a very sharp concentration of light have $\beta^2$ larger than unity. To understand the interaction between nonlinearity and the grain size of phase noise present on the beam, we monitor the variation of $\beta^2$ with propagation distance $z$ over a 15 cm long nonlinear medium for various $P_{\rm in}$ and $L_{\rm coh}$. Figure~\ref{fig:fig4}(a) shows the evolution of $\beta^2$ for a noisy Gaussian beam with $L_{\rm coh}/D_0$ of $0.135$ (blue, solid), 0.075 (red, dot-dashed), 0.045 (green, dashed), and 0.015 (purple, dotted) during linear propagation. The black horizontal dashed line indicates the threshold value of $\beta^2$ above which rogue wave behavior is likely. For a specific set of input parameters ($P_{\rm in}$ and $L_{\rm coh}$), we average $\beta^2$ at each $z$ over 100 different phase masks. This averaged $\beta^2$ is represented by the lines, and the shaded regions around the lines represent its standard deviation. Figures~\ref{fig:fig4}(b) and (c) show the evolution of $\beta^2$ with $z$ for nonlinear propagation with $P_{\rm in}$ of 90 mW and 180 mW, respectively, and the same set of values of $L_{\rm coh}/D_0$ as in Fig.~\ref{fig:fig4}(a).

We note that in all of the scenarios shown in Figs.~\ref{fig:fig4}(a)-(c), $\beta^2$ at first increases with $z$, and then peaks as the phase noise on the beam morphs into intensity distortion. This rate of increase in $\beta^2$ depends strongly on the grain size of the phase noise, as well as the nonlinearity. In the absence of nonlinearity, as observed in Fig.~\ref{fig:fig4}(a), $\beta^2$ peaks when the beam comes to an initial focus along the minima of phase gradients of the added phase mask. When the grain size of the noise is much smaller than the beam diameter (such as when $L_{\rm coh}/D_0 = 0.015$), the phase variations occur over a smaller area within the beam and so the phase gradients are larger and more densely packed [see Fig.~\ref{fig:figC1}(b)]. These grains with large phase gradients within the beam come to a sharp focus after some propagation at which point $\beta^2$ reaches a maximum. For purely linear propagation, these hotspots then diverge, thereby causing $\beta^2$ to decrease with $z$. As the grain size of phase noise becomes larger, the phase gradients decrease in magnitude and become less densely packed [see Fig.~\ref{fig:figC1}(c)], which leads to fewer grains within the beam that focus into hotspots at larger $z$.

In the presence of nonlinearity, the hotspots formed after the initial reorganization of the beam continue to self-focus. Hence, $\beta^2$ increases beyond unity and maximizes when at least one of the hotspots reaches a full width at half maximum (FWHM) size $\Delta r$ of $25 \pm 2.5\mu$m. A Gaussian beam of this FWHM size and an average power of 1.4 mW (say, $P_{\rm cr}$) forms a self-trapped filament that propagates for at least 1.3 cm in the rubidium vapor without any change in its width before diverging due to absorption and diffraction. Filaments of the same width but smaller power than $P_{\rm cr}$ diverge more quickly, while those with power larger than $P_{\rm cr}$ undergo multiple self-focusing and defocusing cycles depending on their power \cite{Feit1988}. For $P_{\rm in}$ of 90 mW and $L_{\rm coh}/D_0$ of 0.015, more than two filaments of size $\Delta r$ are formed when $\beta^2$ is maximized such that the power in each filament is smaller than 0.9 mW [see Fig.~\ref{fig:figC1}(f)]. In contrast, for $P_{\rm in}$ of 90 mW, and $L_{\rm coh}/D_0 \geq 0.045$, a single filament of size $\Delta r$ with average power larger than 1 mW is formed when $\beta^2$ is maximized [see Fig.~\ref{fig:figC1}(k)]. As shown in Fig.~\ref{fig:fig4}(b), this sharper intensity contrast between the ``rogue" filaments and the background intensity in the beam for $L_{\rm coh}/D_0 \geq 0.045$ results in a higher peak of $\beta^2$ for these cases than when $L_{\rm coh}/D_0 \leq 0.045$. When $P_{\rm in}$ is increased to 180 mW, the caustics become even sharper, and more filaments of size $\Delta r$ are formed when $\beta^2$ is maximized, which as shown in Fig.~\ref{fig:fig4}(c) occurs at even smaller $z$ for all cases. For $L_{\rm coh}/D_0$ of 0.015 ($\geq 0.045$), the average power in each filament is smaller (larger) than 1.4 mW [see Fig.~\ref{fig:figC2}]. Hence, for noisy beams with $L_{\rm coh}/D_0 \geq 0.045$, the propagation after the initial peak of $\beta^2$ is followed by another cycle of self-focusing of filaments and subsequently, by diffraction. Nevertheless, even at such large beam powers, the small-grained phase noise seeds the formation of several filaments each containing less than $P_{\rm cr}$ power. This phenomenon limits the maximum intensity in a rogue feature and the tailiness of the intensity statistics.

\section{Conclusion}
In summary, we have shown that the grain size of phase noise on a laser beam can be used to control rogue wave formation in media with a self-focusing nonlinearity. The likelihood of rogue wave formation is minimally affected by nonlinearity when the coherence length of phase noise is much smaller than the beam diameter. Our numerical simulations show that small-grained phase noise causes the beam power to be redistributed into multiple filaments rather than a single filament, which is formed when the phase noise has a longer correlation length. This redistribution of beam power into several filaments of smaller intensity limits the maximum intensity in rogue features relative to the background. Understanding the role of nonlinearity in amplifying the phase noise-induced intensity fluctuations on a field could be helpful in devising efficient mechanisms to mitigate these fluctuations for intense structured light propagating through a turbulent medium \cite{Cox2020, Watkins2020}, and developing efficient radiance limiters using saturable nonlinear media \cite{Schweinsberg2011}. 

\begin{acknowledgments}
This work was supported by the US Office of Naval Research under awards N00014-19-1- 2247 and MURI N00014-20-1-2558. The authors would like to thank J.W. Kuper and G. Marcucci for helpful discussions.
\end{acknowledgments}

\appendix

\section{Susceptibility of rubidium vapor}
\renewcommand{\thefigure}{A\arabic{figure}}
\setcounter{figure}{0}

\begin{figure}[ht!]
	\centering
	\includegraphics[width=0.4\textwidth]{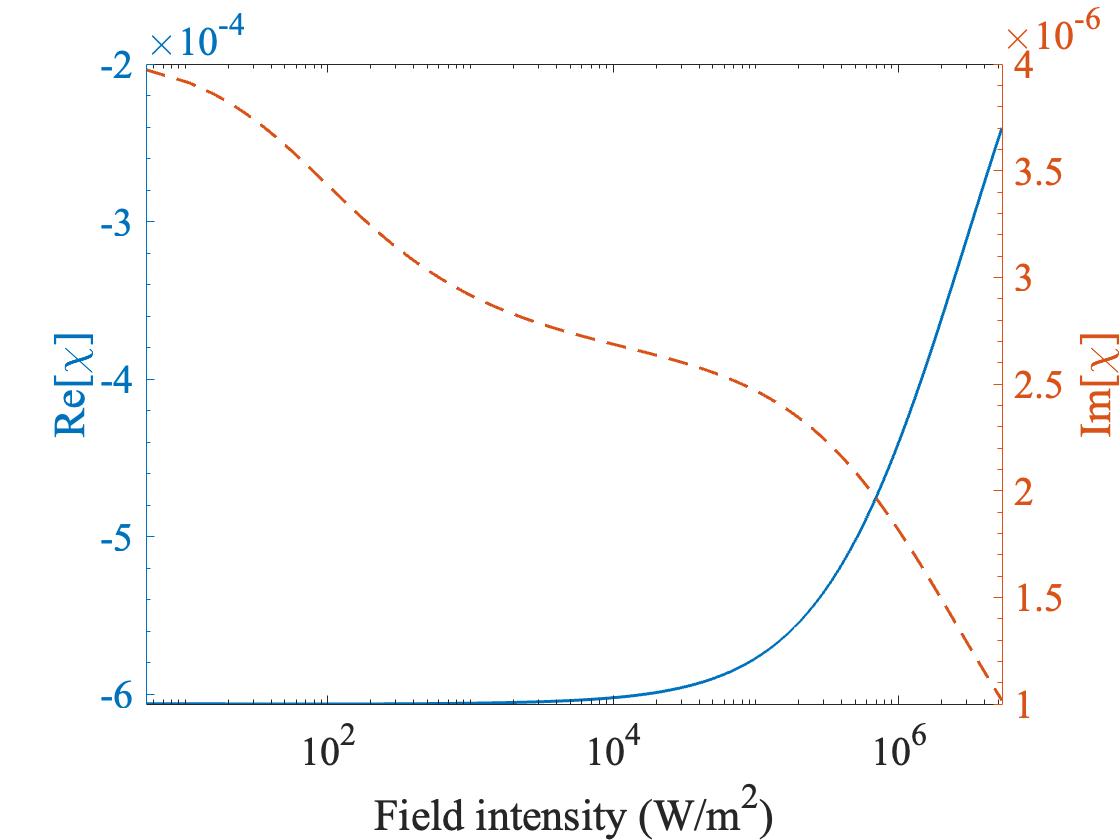}
	\caption{Real (blue, left axis) and imaginary (red, right axis) parts of the total susceptibility of rubidium vapor versus the optical field intensity.}
	\label{fig:figA}
\end{figure}
We use the method described in \cite{Safari2019} to calculate the total susceptibility of rubidium vapor heated to 115 \degree C, and optically pumped at a detuning of 600 MHz above the $^{87}$Rb $D_2$ $F = 1 \rightarrow F' = 2$ transition frequency. We first calculate the susceptibility contribution of each $D_2$ transition of rubidium to the total susceptibility using the equation (6.3.28) in ref. \cite{Boyd_NLO}, and the parameters in ref. \cite{Steck2009}. We include Doppler broadening of the spectrum of each resonant transition by convolving the respective spectrum with the Maxwell distribution of atom velocities \cite{Loudon2000}. We then sum these susceptibility contributions weighted by their oscillator strengths \cite{Siddons2008}. Figure~\ref{fig:figA} shows the real (blue, solid) and imaginary (red, dashed) parts of the total susceptibility $\chi$ of rubidium versus the optical pump intensity.

\section{Power spectral density of the phase noise}
\renewcommand{\thefigure}{B\arabic{figure}}
\setcounter{figure}{0}
\begin{figure}[ht!]
	\centering
	\includegraphics[width=0.45\textwidth]{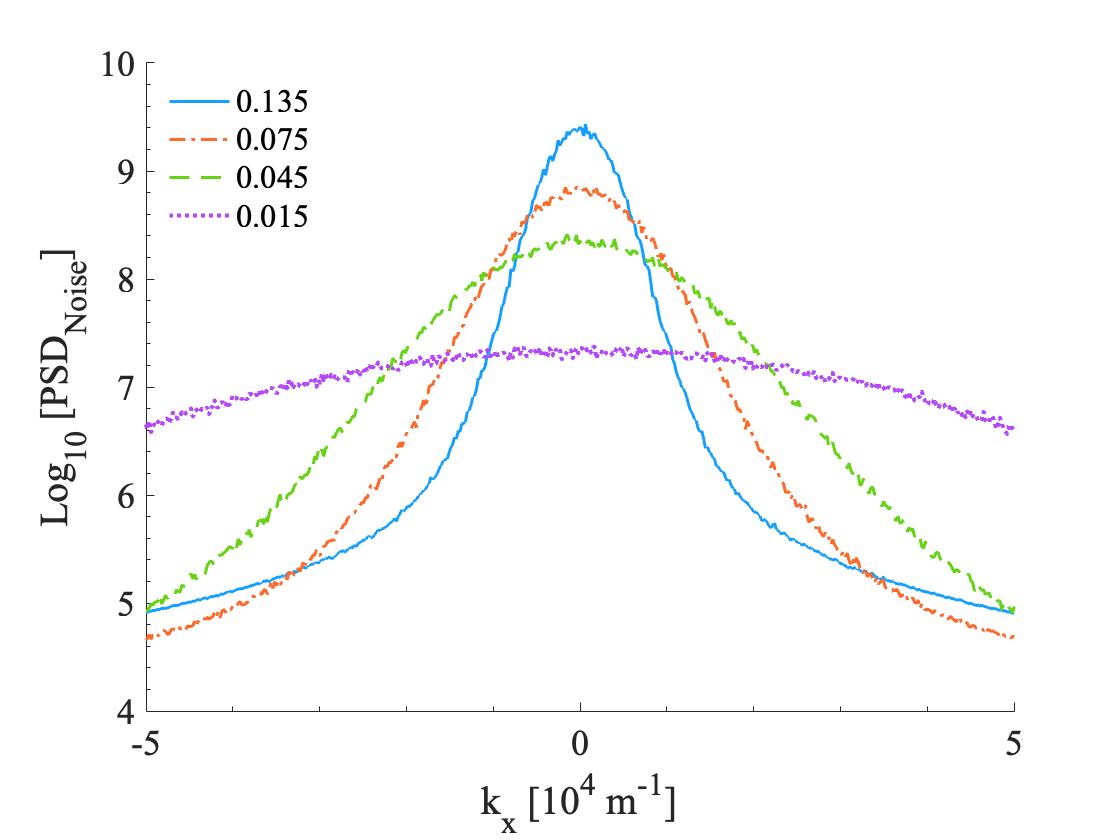}
	\caption{The angular power spectral density (PSD) of phase noise $e^{i\phi_{\text{rand}}(x, y)}$ of various spatial coherence lengths $L_{\rm coh}$. The legend states the values of the corresponding $L_{\rm coh}$ normalized to the beam diameter $D_0$.}
	\label{fig:figB}
\end{figure}

To calculate our phase masks $e^{i\phi_{\text{rand}}(x, y)}$, we first generate a matrix of uniformly distributed random numbers between 0 and 1. We then convolve the matrix with a Gaussian filter, whose response $G(k_x, k_y)$ in the angular frequency space $(k_x, k_y)$ is given by
\begin{equation}
G(k_x, k_y) = \frac{L^2_{\rm coh}}{2\pi} \exp \big[ - \frac{ k^2_x + k^2_y}{2} L^2_{\rm coh} \big],
\label{eq:eqB1}
\end{equation}
with $L_{\rm coh}$ being the coherence length of the phase noise. We then multiply the entire matrix by $\pi$ to rescale the phase variation to be between 0 and $\pi$ rad.

The spectral bandwidth of the phase noise can be estimated from its angular power spectral density (PSD), which we define as the squared magnitude of the 2D Fourier transform of $e^{i\phi_{\text{rand}}(x, y)}$. We take an ensemble average of the PSDs for 250 realizations of phase noise of a particular coherence length $L_{\rm coh}$. In Fig.~\ref{fig:figB}, we show the PSD of phase noise of normalized spatial coherence lengths $L_{\rm coh}/D_0$ of 0.135 (blue, solid), 0.075 (red, dot-dashed), 0.045 (green, dashed) and 0.015 (purple, dotted), with $D_0$ being the Gaussian beam diameter. We note that the PSD of noise becomes more broadband as $L_{\rm coh}/D_0$ is reduced, while the total noise power remains constant.

\section{Field evolution through the cell}
\renewcommand{\thefigure}{C\arabic{figure}}
\setcounter{figure}{0}
\begin{widetext}

\begin{figure}[ht!]
	\centering
	\includegraphics[width=0.95\textwidth]{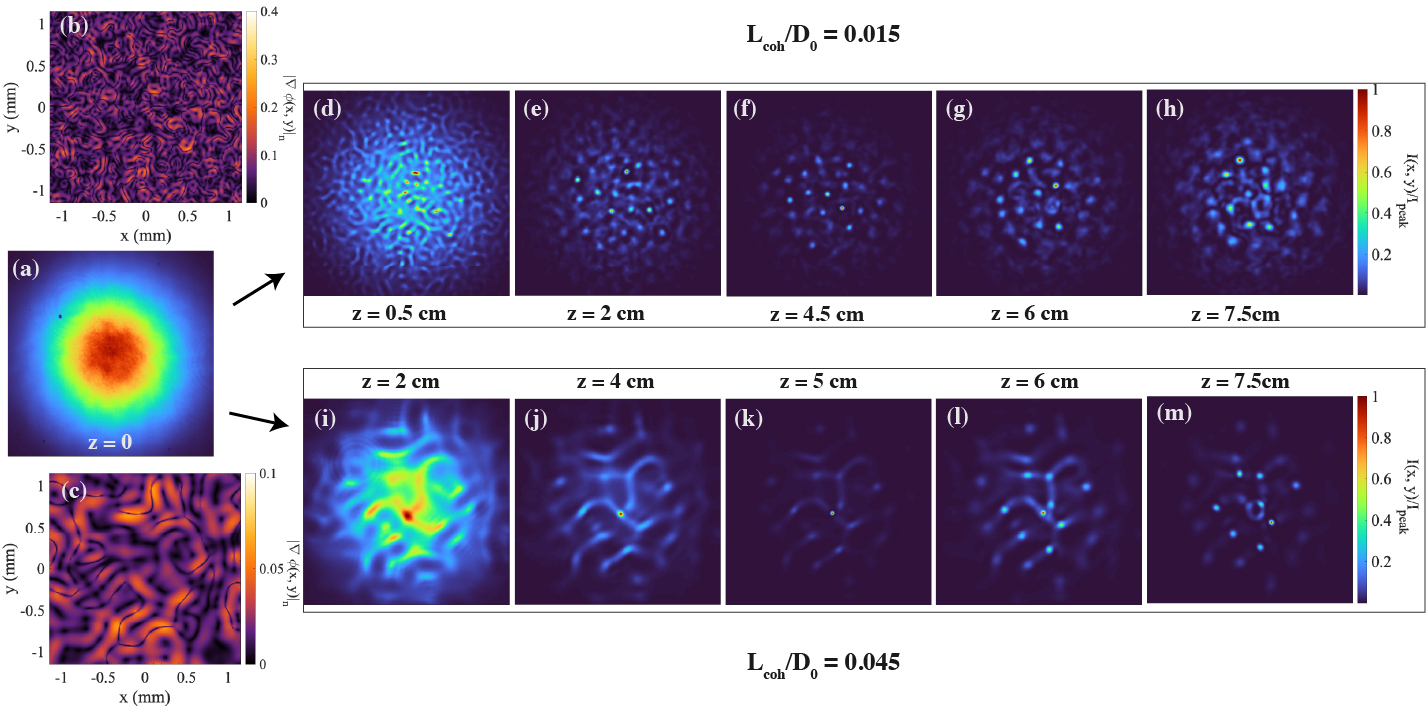}
	\caption{(a) Input Gaussian beam intensity. Phase gradient map $|\boldsymbol{\nabla} \phi(x, y)|$ for a sample mask with $L_{\rm coh}/D_0$ of (b) 0.015 and (c) 0.045. The top panels (d)-(h) show the beam intensity at various propagation distances $z$ within the cell for the phase gradient map shown in (b), and the bottom panels (i)-(m) show the beam intensity at various $z$ for the phase gradient map shown in (c). The beam power $P_{\rm in}$ is 90 mW throughout. The intensity distributions in all panels are normalized with respect to the maximum intensity in the respective frames.}
	\label{fig:figC1}
\end{figure}

\end{widetext}

Figure~\ref{fig:figC1}(a) shows the intensity of the input Gaussian beam generated in our setup. As stated in the main text, the input field intensity in these numerical simulations is taken to be the same as the one generated in the experiment. Figures~\ref{fig:figC1}(b) and (c) show the phase gradient maps $|\boldsymbol{\nabla} \phi(x, y)|$ of a representative random phase mask of coherence lengths $L_{\rm coh}/D_0$ of 0.015 and 0.045, respectively. The top panels (d)-(h) show the normalized intensities of the beam at various propagation distances $z$ stated in the panel label for the phase gradient map shown in Fig.~\ref{fig:figC1}(b). Similarly, the bottom panels (i)-(m) show the normalized intensities of the beam at various $z$ for the phase gradient map shown in Fig.~\ref{fig:figC1}(c). 

\begin{figure}[ht!]
	\centering
	\includegraphics[width=0.49\textwidth]{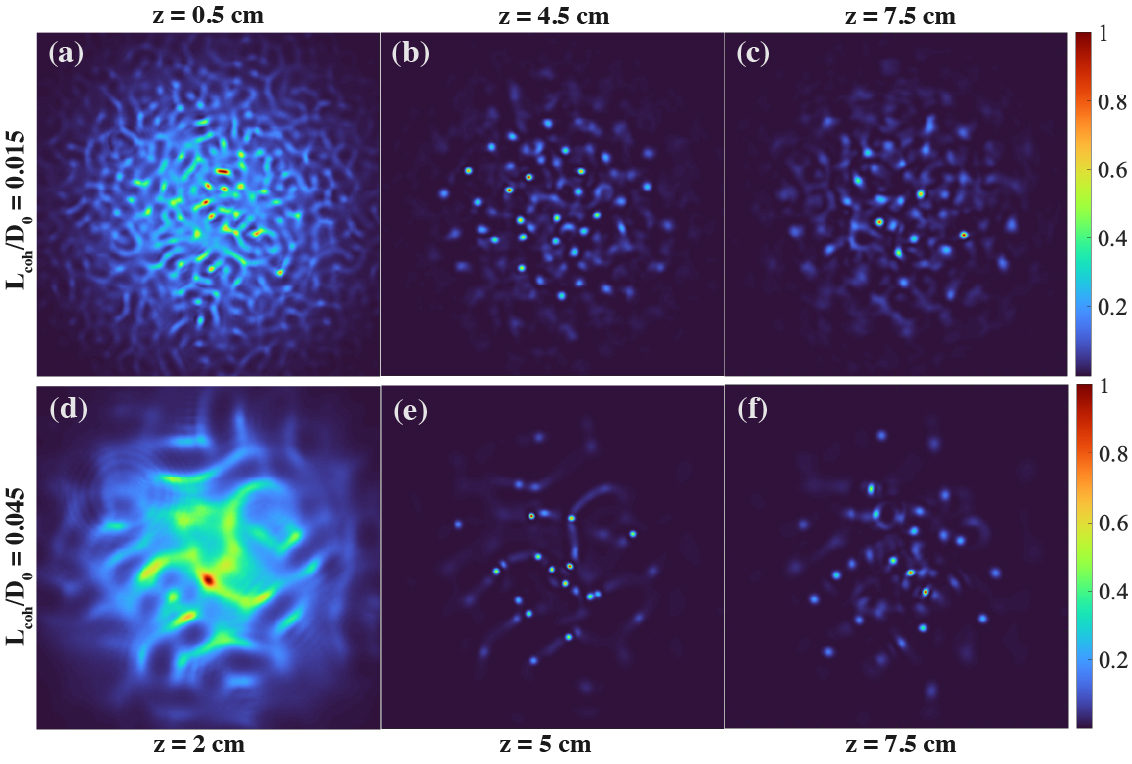}
	\caption{The top panels (a)-(c) show the beam intensity at various propagation distances $z$ within the cell for the phase gradient map shown in Fig.~\ref{fig:figC1}(b), and the bottom panels (d)-(f) show the beam intensity at various $z$ for the phase gradient map shown in Fig.~\ref{fig:figC1}(c). The beam power $P_{\rm in}$ is 180 mW throughout. The intensity distributions in all panels are normalized with respect to the maximum intensity in the respective frames.}
	\label{fig:figC2}
\end{figure}

As shown in Figs.~\ref{fig:figC1}(d) and (i), the beam at first reorganizes by focusing along the minima of the respective phase gradient maps. This initial reorganization occurs at smaller $z$ for phase noise of smaller grain size. The intensity hotspots on this reorganized beam then continue to self focus until at least one of the hotspots reaches the filament width $\Delta r$ as shown in Figs.~\ref{fig:figC1}(f) and (k). The scintillation index of the beam $\beta^2$ is maximized in this plane. The collapse of the filament is limited by absorption, saturation of the nonlinearity, and non-paraxiality \cite{Feit1988}. For $L_{\rm coh}/D_0$ of 0.015, multiple filaments of width $\Delta r$ are formed at $z = 4.5$ cm, and each filament has power smaller than $P_{\rm cr}$ required for forming a self-trapped filament that can propagate for several cm. Hence, these filaments diffract within a few mm as the other hotspots also self-focus and subsequently diffract. Around $z = 6$ cm, absorption losses reduce the effect of nonlinearity, and the filaments start to diverge and $\beta^2$ of the beam starts to decrease with $z$. For $L_{\rm coh}/D_0$ of 0.045, a single filament of width $\Delta r$ and power of 1.1 mW is formed at $z = 5$ cm where $\beta^2$ is also maximized. The large intensity contrast between the filament, and the background intensity of the beam leads to a much larger peak value of $\beta^2$ than the peak value for $L_{\rm coh}/D_0$ of 0.015 even though the power in the filament is still smaller than $P_{\rm cr}$.

Figures~\ref{fig:figC2} (a)-(c) show the beam evolution through the cell for the same phase gradient map as shown in Fig.~\ref{fig:figC1}(b), but at a beam power $P_{\rm in}$ of 180 mW. Similarly, Figs.~\ref{fig:figC2} (d)-(f) show the beam evolution for the phase gradient map as shown in Fig.~\ref{fig:figC1}(c), and at a beam power $P_{\rm in}$ of 180 mW. Comparing Fig.~\ref{fig:figC1}(d) with Fig.~\ref{fig:figC2}(a), and Fig.~\ref{fig:figC1}(e) with Fig.~\ref{fig:figC2}(d), we note that the initial beam reorganization stage involving focusing along the minima of the respective phase gradients remains similar despite the higher power. Comparing Fig.~\ref{fig:figC1}(f) with Fig.~\ref{fig:figC2}(b), and Fig.~\ref{fig:figC1}(k) with Fig.~\ref{fig:figC2}(e), we note that the larger beam power gets distributed into several more filaments along the same underlying caustic pattern.

%


\end{document}